\documentclass[preprint,authoryear,12pt]{elsarticle}
\usepackage[latin1]{inputenc}
\usepackage{amsfonts,amssymb}
\usepackage[english]{babel}
\usepackage{graphicx}
\usepackage{epsfig}
\usepackage{amsfonts,amsmath}
\usepackage{Sweave}
\usepackage{ctable}

\biboptions{compress}

\newtheorem{prop}{Proposition}
\newtheorem{thm}{Theorem}
\newtheorem{lem}[thm]{Lemma}
\newdefinition{rmk}{Remark}
\newproof{pf}{Proof}
\newproof{pot}{Proof of Theorem \ref{thm2}}

\begin{document}

\begin{frontmatter}

\title{Measuring and implementing the bullwhip effect under a generalized demand process}
\author[rvt]{Marlene Silva Marchena\corref{cor1}}
\ead{marchenamarlene@gmail.com, marchena@ele.puc-rio.br}
\address[rvt]{Department of Electrical Engineering, Pontifical Catholic University of Rio de Janeiro.  Rua Marqu\^es de S\~ao Vicente, 225. 
Edificio Cardeal Leme, G\'{a}vea, Rio de Janeiro, Brazil. Cep: 22451-900 Phone: +55 21 35271202 Fax:+55 21 35271232}

\begin{abstract}
The measure of the bullwhip effect, a phenomenon in which demand variability increases as one moves up the supply chain, is a major issue in Supply Chain Management. Although it is simply defined (it is the ratio of the unconditional variance of the order process to that of the demand process), explicit formulas are difficult to obtain. In this paper we investigate the theoretical and practical issues of Zhang [Manufacturing  and Services Operations Management 6-2 (2004b) 195] with the purpose of quantifying the bullwhip effect. Considering a two-stage supply chain, the bullwhip effect is measured for an ARMA(p,q) demand process admitting an infinite moving average representation. As particular cases of this time series model, the AR(p), MA(q), ARMA(1,1), AR(1) and AR(2) are discussed. For some of them, explicit formulas are obtained. We show that for certain types of demand processes, the use of the optimal forecasting procedure that minimizes the mean squared forecasting error leads to significant reduction in the safety stock level. This highlights the potential economic benefits resulting from the use of this time series analysis. Finally, an \textsf{R} function called \texttt{SCperf} is programmed to calculate the bullwhip effect and other supply chain performance variables.  It leads to a simple but powerful tool which could benefit both managers and researchers. 
\end{abstract}


\end{frontmatter}

\section{Introduction}
\label{Introduction}

In recent years, companies in various industries have been able to significantly improve 
their inventory management processes through the integration of information technology into
their forecasting and replenishment systems, and by sharing demand-related information
with their supply chain partners, \cite{Aviv}. However, despite the benefits resulting
from the implementation of the above practices, inefficiencies still persist and are reflected in related costs.

The bullwhip effect, defined as the increase in variability along the supply chain, is
a frequent and expensive phenomenon identified as a key driver of inefficiencies associated with Supply Chain Management (SCM).
It distorts the demand signals, which causes instability in the supply chain, and increases the cost of supplying
end-customer demand. 

\cite{Forrester58} was the first to popularize this phenomenon. Inspired by Forrester's work, several
researchers have studied the bullwhip effect. \cite{Sterman}
used the Beer Game, the most popular simulation of a simple production and distribution system,
to demonstrate that the bullwhip effect is a significant problem with important
managerial consequences. It results in unnecessary costs in supply chains such as inefficient use of production,
distribution and storage capacity, recruitment and training costs, increased inventory and
poor customer service levels (\cite{Metters} and \cite{Lee97b}).

\cite{Lee97a,Lee97b} identified four main causes of the bullwhip effect: demand forecasting, order batching, price fluctuation and supply shortages. 
Of these, demand forecasting is recognised as one of the most important since the inventory system is directly affected by the forecasting technique chosen.
Three popular forecasting methods are commonly used: the Minimum Mean Squared Error (MMSE), Moving Average (MA) and Exponential Smoothing (ES). 

\cite{Chen00a} quantify the bullwhip effect considering the MA forecast method for a simple two-stage supply chain and a first-order autoregressive demand process, AR(1). The authors show that the bullwhip effect is in part due to the effects of demand forecasting. Therefore, given complete access to customer demand information for each stage of the supply chain, the bullwhip effect can be significantly reduced. However, they also show that the bullwhip effect will exist even when demand information is shared by all stages of the supply chain and all stages use the same forecasting technique and the same inventory policy. In similar work \cite{Chen00b} quantify the bullwhip effect considering this time the ES forecast and two different demand processes: AR(1) demand process and a demand
process with a linear trend. In both works, the authors recognize an important limitation of their results: the models considers only non-optimal forecasting methods. The authors justify this limitation saying that ES and MA are commonly used in practice. Users are in general less familiar and less satisfied with more sophisticated methods like time series techniques.

\cite{Zhang04a} investigates the impact of MMSE, MA and ES forecasting methods 
on the bullwhip effect for a simple inventory system in which AR(1)
demand process describes the customer demand and an Order-Up-To (OUT) inventory policy is used. 
The study shows that different forecasting methods lead to bullwhip effect measures with
distinct properties in relation to lead-time and the underlying parameters of the demand
process. The author shows that MMSE forecasting method leads to the lowest
inventory cost. This result is not surprising since MMSE method is optimal when the 
demand model is known to be an AR(1) process. On the other hand, if the
demand structure is not well known, the MA or ES method may
perform better than the MMSE method because they are more flexible.

Another aspect studied in relation of the bullwhip effect is the demand process. A variety of time-series demand models have appeared in the literature of inventory control and SCM. By far, the  AR(1) process is the most frequently
adopted demand model to study the bullwhip effect (\cite{Chen00a,Chen00b}, \cite{Lee97a,Lee97b} and \cite{Zhang04a}).
Recent works use more sophisticated time series models like ARMA and ARIMA \citep{Box} to have more realistic demand models.
\cite{Luong} use an AR(2) and a general AR(p) model; \cite{Duc} use an ARMA(1,1) model. In all these models
an analytical derivation of the bullwhip effect measure is presented and the effects of the autoregressive coefficient on the bullwhip effect is investigated.

\cite{Zhang04b} uses an ARMA(p,q) model to study the demand evolution in supply chains. The author shows that the order history preserves the autoregressive structure of the demand. Zhang's work identifies an important application of this result relating to the quantification of the bullwhip effect. In this paper, inspired by Zhang's work, we study the theoretical and practical issues in order to measure the bullwhip effect for a generalized demand process. In addition, we programmed a function in \textsf{R} \citep{R}, called \texttt{SCperf}\footnote{See the supplementary material}, which implements the bullwhip effect and others supply chain performance variables.
It is well known that measuring the bullwhip effect is difficult in practice but the \texttt{SCperf} function overcomes this 
problem thanks to the help of an \textsf{R} function (\texttt{ARMAtoMA}) which converts an ARMA process into an infinite moving average process. As far as practical applications are concerned, the economic implications of this phenomenon on the inventory cost have been considered. 

Our contributions to this subject can be described as follows: first, this study hopes to improve the understanding of time series techniques. Second, we show that for certain types of demand processes the use of the optimal forecasting procedure that minimizes the mean squared forecasting error leads to significant reduction in the safety stock level. This highlights the potential economic benefits resulting from the use of this time series analysis. Finally, the \texttt{SCperf} function leads to a simple but powerful tool which can be helpful for the study of this phenomenom and other supply chain research problems. 

The structure of our paper is as follows. The next section presents the inventory model. Section \ref{ARMApq case} presents a general ARMA(p,q) case with ARMA(1,1), MA(q), AR(p), AR(1) and AR(2) as particular cases. Next the economic implications are shown. The final section summarizes the main results of the research.

\section{Inventory model}
\label{Inventory model}

In this paper we consider a simple supply chain model for a single item and an OUT 
inventory policy in which the retailer determines a target level or OUT level
and, for every review period, places an order sufficient to bring the inventory position back to this level. 
As did \cite{Chen00b}, we consider that the ordered quantity made in period $t$ is received at the start of period $t+L$ where
$L$ is defined to be a fixed lead time plus the review period, i.e., $L$ is the lead time plus $1$.
For instance, in the case of zero lead time, $L=1$. Shortages are back-ordered and no fixed ordering cost exists.
In the remainder of the paper $L$ will call the lead time. This choise is made for sake of brevity, and should not create confusion.
 
The sequence of events during a replenishment cycle for each period $t$ can be described as follows: the retailer 
receives orders made $L$ periods ago; the demand $d_t$ is observed and satisfied; the retailer
observes the new inventory level and finally places an order $O_t$ to the supplier. As a consequence of this sequence of events, the ordered 
quantity can be written as:
\begin{equation} O_t=S_t-S_{t-1}+d_t,
\quad \label{O}
  \end{equation}
where $S_t$ represents the OUT level in period $t$, i.e., the inventory position at the beginning of period
$t$. Note that in the above expression, we have implicitly assumed that the order quantity can be negative, i.e., returning items are allowed at no costs.
This unpleasant feature is needed for tractability. However, the free-return assumption 
becomes negligible when the demand mean is sufficiently large. Further detail
about this assumption can be found in \cite{Lee00} and \cite{ChenLee}.

Under the OUT policy, the OUT level $S_t$ can be estimated from the observed demand as:

\begin{equation} S_t=\hat{D}^L_t+z\hat{\sigma}^L_t,
\quad \label{order_up}
  \end{equation}
where $\hat{D}^L_t=\sum_{\tau=1}^{L}\hat{d}_{t+\tau}$ is an estimate of the mean demand over $L$ periods after period
$t$, $z$ is the safety factor which is a fixed constant chosen 
to meet a required service level and $\hat{\sigma}^L_t=\sqrt{Var(D_t^L-\hat{D}_t^L)}$ is an estimate of
the standard deviation of $L$ periods forecast error. An OUT policy of this form is optimal when the 
demand came from a normal distribution and there is no setup or fixed
order cost. 

As \cite{Chen00b} mention, if the retailer follows an OUT policy of the form $S_t=D^L+z\sigma^L$,
where $D^L$ is the known mean and $\sigma^L$ is the standard deviation of the demand over $L$ periods, then
the OUT level in any period is constant and, consequently, the order is equal to the last observed
demand. Therefore, there is no bullwhip effect. However, these
values are, in general, unknown and the retailer must estimate them using some forecasting 
technique. Note that the introduction of forecasting values in the calculation of $S_t$ is one of the main 
causes for the variability increase along the supply chain or, in other words, the bullwhip effect.

The demand forecast is performed here by using the MMSE method. It was 
shown that, for an ARMA process, the MMSE forecast for period $t+\tau$ is the conditional mean given the observed information\footnote{Box and Jenkins, 1970, pp.128.}.
Let $\digamma_t=\{d_t,d_{t-1}, ....\}$ be the information set which represents
all the information available until period $t$. Hence, the demand forecast for $\tau$ periods ahead is given by $E(d_{t+\tau}|\digamma_t)$. 

In order to quantify the bullwhip effect we combine (\ref{O}) and (\ref{order_up}) to rewrite the order quantity as:

\begin{equation} O_t=(\hat{D}^L_t-\hat{D}^L_{t-1})+z(\hat{\sigma}^L_t-\hat{\sigma}^L_{t-1})+d_t.
\quad \label{order}
  \end{equation}
We show later in the paper (see Lemma \ref{lem1}) that the standard deviation of lead-time forecast error remains constant over time for an ARMA(p,q) demand process. Hence, $\hat{\sigma}^L_t=\hat{\sigma}^L_{t-1}$ and the order quantity given in (\ref{order}) becomes

\begin{equation} O_t=(\hat{D}^L_t-\hat{D}^L_{t-1})+d_t.
\quad \label{order1}
  \end{equation}

Let $M$ be the measure for the bullwhip effect. Since $M$ can be obtained from 
the ratio between the unconditional variance of the order process to that of the demand process, we have

\begin{equation} M=\frac{Var(O_t)}{Var(d_t)}.
\quad \label{bullwhip}
  \end{equation}
Note that $M$ is calculated by using the variances from both side of Equation (\ref{order1}). The fact that
$M=1$ means that there is no variance amplification, while $M>1$ means that the bullwhip effect is present. On the other hand,
$M<1$ means that the orders are smoothed if compared with the demand. The last case is less common since it is unlikely to have a situation where stages up the supply chain have a better representation of the customer demand than the first stage (i.e., the retailer). 

In what follows, the corresponding bullwhip effect measure is derived for a general ARMA(p,q) demand
process and some particular cases are discussed. Since the calculation is complex, we cannot always express
this measure in a closed form. In this context, the \texttt{SCperf} function was developed to overcome this computational difficulty.

\section{ARMA(p,q) case}
\label{ARMApq case}

The demand process, $d_t$, seen by the retailer, is described by a stationary ARMA(p,q) process 
as follows\footnote{Our representation differs from some works where the MA model is written
with negative coefficients, i.e., $d_t=\mu+\phi_1{d_{t-1}}+\cdots+\phi_p{d_{t-p}}+\epsilon_t-\theta_1\epsilon_{t-1}-
\cdots-\theta_q\epsilon_{t-q}$. We chose this representation to be in accordance with the \textsf{R}
software which was used to implement the bullwhip effect.}:

\begin{equation}d_t=\mu+\phi_1{d_{t-1}}+\cdots+\phi_p{d_{t-p}}+\epsilon_t+\theta_1\epsilon_{t-1}+\cdots+\theta_q\epsilon_{t-q},
\quad \label{armapq}
  \end{equation}
where $\mu$ is a nonnegative constant, $\epsilon_t$ is i.i.d. normally distributed, with mean zero and variance
$\sigma_\epsilon^2$, $p$ is the autoregressive order of the process, $q$ is the moving average order of the process, $\phi_j$ is the autoregressive coefficient, and $\theta_j$ denotes the moving average coefficient. It is often useful to express (\ref{armapq}) in terms of
the lag operator, B, where $B^kd_t = d_{t-k}$. In order to do so, let $\phi(B)=1-\phi_1B-\cdots-\phi_pB^p$ and $\theta(B)=1+\theta_1B+\cdots+\theta_qB^q$. Hence, the demand process in (\ref{armapq}) can be expressed as:
$$\phi(B)d_t=\mu+\theta(B)\epsilon_t,$$
where $\phi(B)$ and $\theta(B)$ are known as the autoregressive and the moving average polynomials in the lag operator of degree $p$ and $q$.
If we substitute the lag operator by a constant $z$, we get the characteristic equations:
$$\phi(z)=1-\phi_1z-\phi_2z^2-\cdots-\phi_pz^p$$
and
$$\theta(z)=1+\theta_1z+\theta_2z^2+\cdots+\theta_qz^q.$$
The process is called the autoregressive process of order $p$,
AR(p), if $\theta(z)=1$ and a moving average process of order $q$, MA(q), if $\phi(z)=1$. We assume that the process described in (\ref{armapq}) is invertible and covariance stationary, i.e., the roots of the equations $\theta(z)=0$ and $\phi(z)=0$ must be outside the unit circle. To avoid the problem of parameter redundancy, it is assumed that the two characteristic equations share no common roots. 

It is important to note that the constant $z$ in the above equations is different from the constant used to define the safety factor. We have chosen this notation to be in accordance with time series notation and we hope that this will not cause any future confusion. Using stationarity and taken expectations in (\ref{armapq}) directly it can be found that the mean of ARMA(p,q) demand process is defined by

\begin{equation}\mu_d=\frac{\mu}{1-\phi_1-\cdots-\phi_p}.
\quad \label{mean}
  \end{equation}
It is known from time series theory that a stationary ARMA(p,q) demand process under the above conditions 
can be written as an infinite moving average process of its errors, $MA(\infty)$, that is,

\begin{equation} d_t=\mu_d+\Sigma_{j=0}^\infty\psi_j\epsilon_{t-j},
 \quad \label{MAinf}
  \end{equation}
where $\mu_d$ is defined as in Equation (\ref{mean}) and the sequence $\{\psi_j\}$ in (\ref{MAinf}) is determined by the relation 
$\psi(z)=\sum_{j=0}^{\infty}\psi_jz^j=\frac{\theta(z)}{\phi(z)}$, or equivalently by the identity
$$(\psi_0+\psi_1z+\psi_2z^2+\cdots)(1-\phi_1z-\phi_2z^2-\cdots-\phi_pz^p)=(1+\theta_1z+\theta_2z^2+\cdots+\theta_qz^q).$$
Equating coefficients of $z^j$, $j=0, 1, . . .,$ we find that

\begin{equation} \psi_j=\sum_{k=1}^{p}\phi_k\psi_{j-k}+\theta_j \mbox{ for } j \geq 1,
 \quad \label{particular_psi}
  \end{equation}
where $\theta_0=1$, $\theta_j=0$ for $j > q$, and $\psi_j=0$ for $j<0$. Note that equation (\ref{particular_psi}) is a recursive equation.
Therefore, the $\psi$-weights satisfy the homogeneous difference equation given by
\begin{equation} \psi_j-\sum_{k=1}^{p}\phi_k\psi_{j-k}=0, \mbox{   }j\geq max(p, q + 1),
 \quad \label{general_psi}
  \end{equation}
with initial conditions given by equation (\ref{particular_psi}). From homogeneous difference equation 
theory the general solution for equation (\ref{general_psi}) can be read off directly as:
\begin{equation}\psi_j =c_1z_1^{-j} +\cdots+c_rz_p^{-j},
\quad \label{sol_psi}
  \end{equation}
where $z_1,..,z_p$ are distinct roots of the polynomial $\phi(z)$ and $c_k$, for $k=1, 2, . . . , p$ are constants which depend on the initial 
conditions.\footnote{In the case of the repeated root, the solution is different. See \cite{Shumway} for a brief and heuristic account of the topic. For 
details about homogeneous difference equation theory the reader is referred to \cite{Mickens87}.} Now, from equation (\ref{MAinf}), the variance of the demand process can be expressed as:
\begin{equation} \sigma_d^2=\sigma^2_\epsilon\sum_{j=0}^\infty\psi^2_j.
\quad \label{vardemand}
  \end{equation}
  
It is important to note that the $MA(\infty)$ representation depends on an infinite number of parameters and, 
consequently, it is not directly useful in practical applications. On the other hand, \cite{Zhang04b}, using the $MA(\infty)$ representation, shows a property, 
called by the author ARMA-in-ARMA-out (AIAO), which reveals that the order history preserves the autoregressive structure of the demand
and transforms its moving average structure according to a simple algorithm\footnote{\citealt[pp. 197]{Zhang04b}}. As the author remarks, the practical value of the AIAO property lies in its ability to make simpler the measuring of the bullwhip effect. 

\begin{prop}\citep{Zhang04b} The retailer's demand process can be represented by an $MA(\infty)$ process with 
respect to the retailer's full information shocks $\epsilon_t$, as in equation (\ref{MAinf}). Hence, the retailer's order $O_t$ to its supplier is
given by:

\begin{equation}
O_t=\mu_d+\sum_{j=0}^L\psi_j\epsilon_t+\sum_{j=1}^\infty\psi_{L+j}\epsilon_{t-j}
 \quad \label{prop1}
  \end{equation}
where the $\psi_j=0$ for $j<0$, $\psi_0=1$, and $\psi_j=\sum_{k=1}^{p}\phi_k\psi_{j-k}+\theta_j$ for $j \geq 1$.
\end{prop}

\pf See \cite{Zhang04b}. \hfill $\square$

\begin{prop} For a stationary ARMA(p,q) demand process, the measure for the bullwhip effect is defined by:

\begin{equation} M=1+\frac{2\sum_{i=0}^L\sum_{j=i+1}^L\psi_i\psi_j}{\sum_{j=0}^\infty\psi_j^2},
 \quad \label{BE}
  \end{equation}
where the $\psi_j=0$ for $j<0$, $\psi_0=1$, and $\psi_j=\sum_{k=1}^{p}\phi_k\psi_{j-k}+\theta_j$ for $j \geq 1$.
\end{prop}

\pf Taking the variance of the order quantity, Equation (\ref{prop1}), we have
$Var(O_t)=\sigma_\epsilon^2(\sum_{j=0}^L\psi_j)^2+\sigma_\epsilon^2\sum_{j=1}^\infty\psi_{L+j}^2
=\sigma_\epsilon^2(\sum_{j=0}^\infty\psi_j^2+2 \sum_{i=0}^L \sum_{j=i+1}^L\psi_i\psi_j)$. We complete the proof by substituting this result and (\ref{vardemand}) in (\ref{bullwhip}).   \hfill $\square$

\begin{prop} The bullwhip effect increases when the lead-time $L$ increases if and only if $\psi_{L+1}\sum_{j=0}^{L}\psi_j>0$.

\end{prop}

\pf From equation (\ref{BE}), it is straightforward to see that the bullwhip effect exists, i.e., $M>1$, if and only if  $\sum_{i=0}^L\sum_{j=i+1}^L\psi_i\psi_j>0$.
Let $g(L)=\sum_{i=0}^L\sum_{j=i+1}^L\psi_i\psi_j$ and $\triangle g(L)=g(L+1)-g(L)$. Then 

\noindent $\triangle g(L)=\sum_{i=0}^{L+1}\sum_{j=i+1}^{L+1}\psi_i\psi_j-\sum_{i=0}^L\sum_{j=i+1}^L\psi_i\psi_j
=\psi_0(\sum_{j=1}^{L+1}\psi_j-\sum_{j=1}^L\psi_j)+\cdots+\psi_{L-1}(\sum_{j=L}^{L+1}\psi_j-\psi_L)+\psi_L\psi_{L+1}
=\psi_{L+1}\sum_{j=0}^L\psi_j$. Hence, $\triangle g(L)>0$ if and only if $\psi_{L+1}\sum_{j=0}^L\psi_j>0$. Hence, $g(L)$ is a non-decreasing function of the lead-time $L$ if and only if $\psi_{L+1}\sum_{j=0}^L\psi_j>0$.  \hfill $\square$

\subsection{ARMA(1,1) case}

The stationary ARMA(1,1) demand process is described as follow:

\begin{equation}
d_t=\mu+\phi{d_{t-1}}+\epsilon_t+\theta\epsilon_{t-1}.
\quad \label{arma11}
  \end{equation}
Stationarity and invertible conditions impose $|\phi|<1$
and $|\theta|<1$. It can be shown that the mean and variance of the demand process are
$\mu_d=\frac{\mu}{1-\phi_1}$ and $\sigma_d^2=\frac{(1+\theta^2+2\phi\theta)\sigma_\epsilon^2}{1-\phi^2}$, respectively. 
 
\begin{prop} For a stationary ARMA(1,1) demand process the measure for the bullwhip effect is defined by:

\begin{equation}
M(L,\phi,\theta)=1+\frac{2(\phi+\theta)(1-\phi^L)}{(1-\phi)(1+\theta^2+2\phi\theta)}\left[1-\phi^{L+1}+\theta\phi(1-\phi^{L-1})\right].
 \quad \label{bullwhip_arma11}
  \end{equation}
\end{prop}

\pf Since the AR polynomial associated with (\ref{arma11}) is $\phi(z)=1-\phi z$, and its root, say $z_1$, is $z_1=\phi^{-1}$,
then the general solution for the $\psi$-weights can be written directly from equation (\ref{sol_psi}) as $\psi_j =c\phi^j$. From (\ref{particular_psi}) 
we find that the initial conditions are $\psi_0=1$ and $\psi_1= \phi+\theta$, which
combining with the general solution, results in $c=(\phi+\theta)/\phi$. Hence, $\psi_j =(\phi+\theta)\phi^{j-1}$ for $j\geq 1$.
Since we know $\psi_j$, we can rewrite the follow relations as:
\begin{eqnarray*}
\sum_{i=0}^L\sum_{j=i+1}^L\psi_i\psi_j&=&\psi_0\sum_{j=1}^L\psi_j+\sum_{i=1}^L\sum_{j=i+1}^L\psi_i\psi_j\\
&=&(\phi+\theta)\frac{1-\phi^L}{1-\phi}+\frac{\phi(\phi+\theta)^2(1-\phi^L)(1-\phi^{L-1})}{(1-\phi)(1-\phi^2)}\\
&=&\frac{(\phi+\theta)(1-\phi^L)}{(1-\phi)(1-\phi^2)}\left[1-\phi^{L+1}+\theta\phi(1-\phi^{L-1})
\right]
\end{eqnarray*}
and
$$\sum_{j=0}^\infty\psi^2_j=\frac{1+\theta^2+2\phi\theta}{1-\phi^2}.$$
Substituting the two above results in equation (\ref{BE}) we complete the proof.  \hfill $\square$
Using a generalized formula for the variance ratio, we get a similar expression to that obtained by \cite{Duc}.
There are two other results found by the above authors which are easily verified. 

\begin{figure}[t]
 \begin{center}
 \includegraphics[scale=0.1]{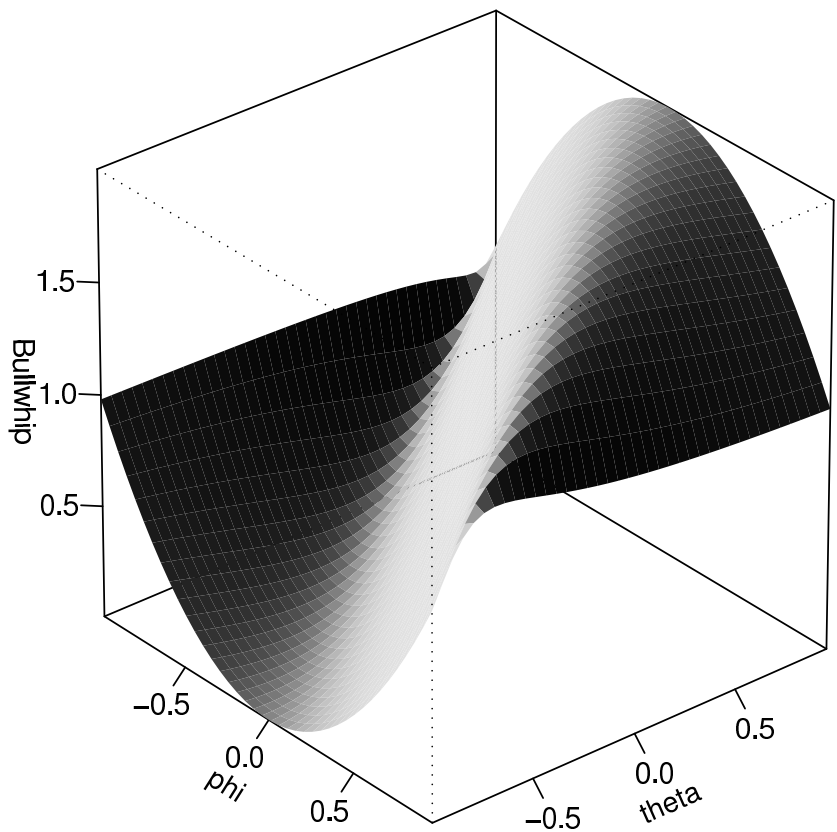}
 \caption{Bullwhip generated with ARMA(1,1) demand process when L=1}\label{fig1}
 \end{center}
  \end{figure}

\bigskip

\begin{prop} The bullwhip effect exists, i.e, $M(L,\phi,\theta)>1$, if and only if, $\phi+\theta>0$.
\end{prop}

\pf \citealt[pp. 248-249]{Duc}. \hfill $\square$

\begin{prop} The bullwhip effect, measured by $M(L,\phi,\theta)$, has the following properties.

(a) If $\phi>0$, the bullwhip effect increases as $L$ increases.

(b) If $-\theta<\phi<0$ and $L$ is an odd number, the larger $L$ is, the smaller the bullwhip effect is.

(c) If $-\theta<\phi<0$ and $L$ is an even number, the larger $L$ is, the larger the bullwhip effect is.
\end{prop} 

\pf \citealt[pp. 249]{Duc}. \hfill $\square$
\medskip

\begin{figure}[t]
 \begin{center} \includegraphics[scale=0.4]{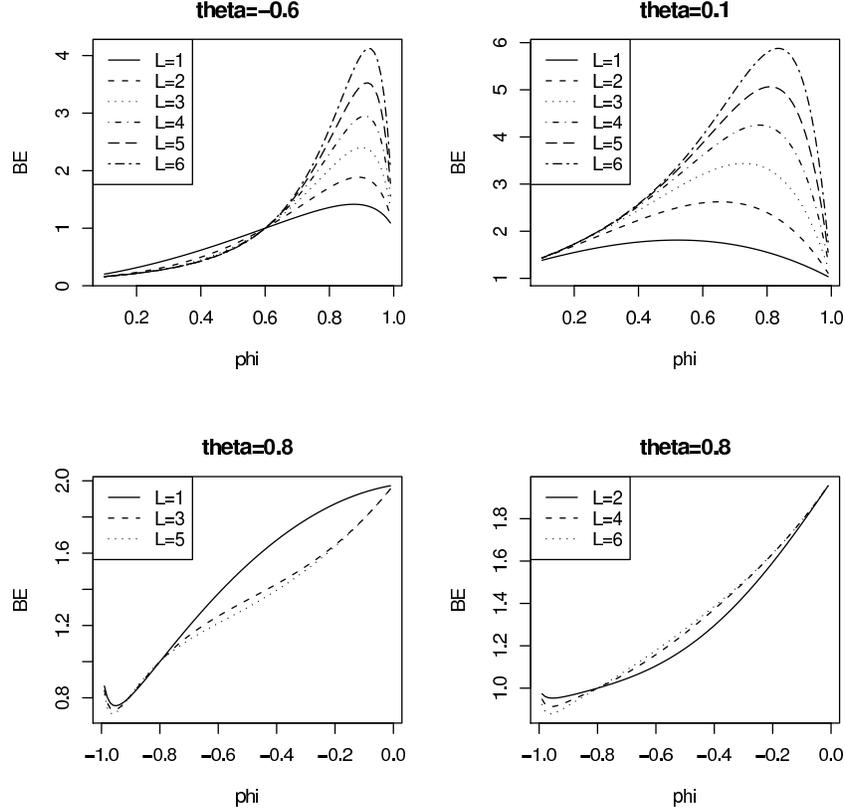}
 \caption{Effect of the AR coefficient on BE for different values of theta}\label{fig2}
 \end{center}
 \end{figure}

In conclusion the bullwhip effect occurs only when the sum of the AR parameter and the MA parameter is larger than zero ( See Figure~\ref{fig1}) and it does not always increase when the lead time $L$ increases. In fact, if $\phi+\theta>0$ and $\phi>0$ the bullwhip effect increases when the lead-time increase. However, if $-\theta<\phi<0$ and $L$ is an odd number, the bullwhip effect becomes smaller as $L$ becomes larger; if $-\theta<\phi<0$ and $L$ is an even number, the bullwhip effect becomes larger as $L$ becomes larger. Figure~\ref{fig2} represents situations where these facts are observed.

\subsection{MA(q) case}
The MA(q) demand process can be written as

$$d_t=\mu+\sum_{j=0}^{q}\theta_j\epsilon_{t-j}=\mu+(1+\theta_1B+\cdots+\theta_qB^q)\epsilon_t=\mu+\theta(B)\epsilon_t.$$
Since $\theta(B)$ is finite, no restrictions on
the $MA$ parameters are needed to ensure stationarity. Considering $q \rightarrow \infty$ the
infinite $MA$ representation is written as:
$$d_t=\mu_d+\sum_{j=0}^{\infty}\psi_j\epsilon_{t-j},$$
where $\psi_j=\theta_j$ for $j=0,1,..,q$ and $\psi_j=0$ for $j>q$. It 
can be easily seen that $\mu_d=\mu$ and $\sigma_d^2=(1+\theta_1^2+\cdots+\theta_q^2)\sigma_\epsilon^2$.
Since the above demand process is i.i.d. the OUT level, $S_t$, is constant across all periods. Hence, 
from Equation (\ref{O}), $O_t=d_t$, consequently, the bullwhip ratio equals one. 

\subsection{AR(p) case}
The stationary AR(p) demand process is described as follow:
\begin{equation*}
d_t=\mu+\phi_1{d_{t-1}}+\cdots+\phi_p{d_{t-p}}+\epsilon_t
\end{equation*}
Assume that the AR parameters are such that $\{d_t\}$ is stationary. It is straightforward to verify that the $MA(\infty)$
representation is
$$d_t=\mu_d+\psi(B)\epsilon_t,$$
where $\mu_d$ is defined as in (\ref{mean}) and
$\psi(B)=\phi^{-1}(B)$. The $\psi$-weights in the $MA(\infty)$ representation
of $d_t$ are found directly from (\ref{sol_psi}) and it can be shown that the constants
are expressed by:
\begin{equation} c_i =\frac{z_i^{p-1}}{\prod_{k=1 k\neq i}^p (z_i-z_k)},
\quad \label{sol_c}
  \end{equation}
where the constants terms $c_i$ sum to the unity,
$c_1+\cdots+c_p=1$, see \citealt[pp. 33-36,]{Hamilton} for details.

\subsection{AR(1) case}
The stationary AR(1) demand process is described as follows:

\begin{equation}
d_t=\mu+\phi{d_{t-1}}+\epsilon_t.
 \quad \label{ar1}
  \end{equation}
Stationarity condition imposes $|\phi|<1$. Using stationarity it can be shown that
the mean and the variance of the process are $\mu_d=\frac{\mu}{1-\phi_1}$ and
$\sigma^2_d=\frac{\sigma_\epsilon^2}{1-\phi_1^2}$, respectively. 

\begin{prop} For a stationary AR(1) demand process the measure for the bullwhip effect is defined by:

\begin{equation} M(L,\phi)=1+\frac{2\phi(1-\phi^L)(1-\phi^{L+1})}{1-\phi}
 \quad \label{bullwhip_ar1}
  \end{equation}
\end{prop}

\pf As in the ARMA(1,1) case, the AR polynomial associated with 
(\ref{ar1}) is $\phi(z)=1-\phi z$, and the root, say,
$z_1$, is $z_1=\phi^{-1}$. Using (\ref{sol_psi}) the general
solution is $\psi_j=c(z_1)^{-j}= c\phi_1^j$ with $\psi_0=1$ and $\psi_1 =\phi$ as 
initial conditions. Combining the general solution with the initial conditions we find 
$\psi_j=\phi^j$. Since $\psi_j=\phi^j$, Equation (\ref{BE}) can be expressed as:

\begin{equation}
M(L,\phi)=1+\frac{2\sum_{i=0}^L\sum_{j=i+1}^L\phi^i\phi^j}{\sum_{j=0}^\infty\phi^{2j}},
\quad \label{BE_ar1}
\end{equation}
where 
\begin{eqnarray*}
\sum_{i=0}^L\sum_{j=i+1}^L\phi^i\phi^j&=&\sum_{i=0}^L\sum_{k=0}^{L-i-1}\phi^i\phi^{k+i+1}=\frac{\phi}{1-\phi}\sum_{i=0}^L\phi^{2i}(1-\phi^{L-i})\\
&=&\frac{\phi}{1-\phi} \left[\frac{1-\phi^{2(L+1)}}{1-\phi^2}-\frac{\phi^L(1-\phi^{L+1})}{1-\phi}\right]\\
&=&\frac{\phi}{1-\phi}\left[\frac{(1-\phi^L)(1-\phi^{L+1})}{1-\phi^2} \right]
\end{eqnarray*}
and $\sum_{j=0}^\infty\phi^{2j}=\frac{1}{1-\phi^2}$.
Substituting the two above results in (\ref{BE_ar1}) complete the proof. \hfill $\square$
 
\begin{figure}[t]
 \begin{center}
 \includegraphics[scale=0.4]{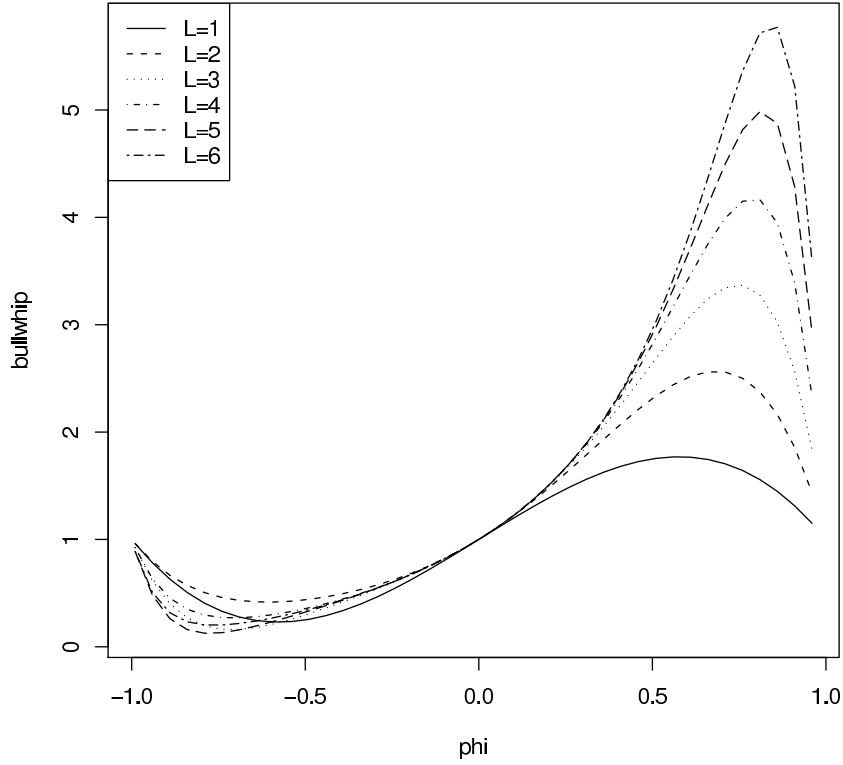}
 \caption{Relationship between the bullwhip effect and demand autocorrelation} \label{fig3}
 \end{center}
\end{figure}

\begin{prop} For a stationary AR(1) demand process the bullwhip effect, measured by Equation (\ref{bullwhip_ar1}), 
has the following properties:

(a) The bullwhip effect exists, i.e, $M(L,\phi)>1$, if and only if $\phi>0$.

(b) For $\phi > 0$, a longer lead-time leads to a more significant bullwhip effect.
\end{prop}

\pf Since $1-\phi>0$, $1-\phi^L>0$ and $1-\phi^{L+1}>0$ for $|\phi|<1$, it is 
straightforward to see that $M(L,\phi)>1$, if and only if $\phi>0$. Let $f(L,\phi)=\phi(1-\phi^L)(1-\phi^{L+1})$ 
and $\triangle f(L)\equiv f(L+1,\phi)-f(L,\phi)$. Then, $\triangle f(L)=(1-\phi^2)(1-\phi^{L+1})\phi^{L+1}$.
It can be easily seen that $\triangle f(L)$ is an increasing function with respect to $L$ since $\phi>0$. Hence, the bullwhip effect, i.e, 
$M(L,\phi)$, increases as $L$ increases since $\phi>0$. \hfill $\square$
\medskip

\noindent Figure \ref{fig3} depicts how the bullwhip effect generated by AR(1) demand process increases for different lead-time values, $L=1,...,6$.
We can observe that the increase of the lead-time has a strong impact on the bullwhip effect when $\phi>0.5$ and a less significant one when $\phi$ is positive and near zero and one. Therefore, as it was already noted by \cite{Zhang04a}, reduction on the lead-time can reduce the bullwhip effect if the demand autocorrelation is positive and away from zero and unity in the case of AR(1) demand process. 

\subsection{AR(2) case}

The stationary AR(2) demand process satisfies:

\begin{equation}
d_t=\mu+\phi_1{d_{t-1}}+\phi_2{d_{t-2}}+\epsilon_t
\quad \label{BE_ar2}
\end{equation}
In the AR(2) case, stationarity implies that the roots
of $\phi(z)=0$ lie outside the unit circle or, equivalently,
the parameters $\phi_1$ and $\phi_2$ must lie in the triangular
region restricted by $\phi_1+\phi_2<1$, $\phi_2-\phi_1<1$ and
$|\phi_2|<1$. It can be shown that for a stationary $AR(2)$ demand process the mean and variance of the demand are 
$\frac{\mu}{1-\phi_1-\phi_2}$ and $\frac{(1-\phi_2)\sigma_\epsilon^2}{(1+\phi_2)[(1-\phi_2)^2-\phi_1^2]}$, respectively.

\begin{prop} Let $z_1$ and $z_2$ be the solutions for the characteristic equation defined by the AR(2) process. For
a stationary AR(2) demand process the $\psi$-weights are defined by:

$$\psi_j =\frac{z_2^{1+j}-z_1^{1+j}}{z_1z_2(z_2-z_1)}$$
\end{prop}

\pf From Equation (\ref{general_psi}), the general solution for $\psi_j$-weights
for an AR(2) process is described by:

\begin{equation} \psi_j =c_1(z_1)^{-j}+c_2(z_2)^{-j}
 \quad \label{sol_psiAR2}
  \end{equation}
where 
\begin{equation} z_1=\frac{-\phi_1+\sqrt{\phi_1^2+4\phi_2}}{2\phi_2},
 \quad \label{z1}
  \end{equation}
and
\begin{equation} z_2=\frac{-\phi_1-\sqrt{\phi_1^2+4\phi_2}}{2\phi_2}
 \quad \label{z2}
  \end{equation}
are the solutions for the characteristic equation
$1-\phi_1z-\phi_2z^2=0$. On the other hand, from Equation (\ref{sol_c}), the values of the constants are given by:
\begin{equation} c_1=\frac{z_1^{-1}}{z_1^{-1}-z_2^{-1}}
 \quad \label{c1}
  \end{equation}
and
\begin{equation} c_2=-\frac{z_2^{-1}}{z_1^{-1}-z_2^{-1}}
 \quad \label{c2}
  \end{equation}
Finally by replacing (\ref{z1}), (\ref{z2}), (\ref{c1}) and (\ref{c2}) in (\ref{sol_psiAR2}) we find the result. \hfill $\square$
\medskip

\noindent Note that the solution for the $\psi_j$-weights are a
function of the roots of the AR polynomial. In the AR(2) case,
the roots can be real if $\phi_1^2+4\phi_2>0$, or complex if
$\phi_1^2+4\phi_2<0$. In both cases, from a computational point of
view, the solution for the $\psi_j$-weights can be found and,
therefore, we can get a measure for the bullwhip effect. Since an explicit form 
for the measure for the bullwhip effect is difficult to obtain, we investigated the relation of the 
autoregressive coefficients and lead-time by numerical experimentation. For 
an analytical derivation the reader is referred to \cite{Luong}.

When $\phi_1<0$, the bullwhip effect does not exist for $\phi_2\leq 0$ and for $\phi_2>0$, $\phi_2-\phi_1<1$.
On the other hand, when $\phi_1>0$ the bullwhip effect always exists for $\phi_2>0$, $\phi_1+\phi_2<1$ and for
$\phi_2<0$, $\phi_1+\phi_2<1$. The pattern shown when the lead-time is equal to one does not seem to be the same when the lead-time increases. Using the function \texttt{SCperf}, it can be verified that the there is no bullwhip effect when $\phi_1<0$ and $\phi_2\leq 0$ and always does when $\phi_1>0$, 
$\phi_2>0$ and $\phi_1+\phi_2<1$. In the last case, we observe that the bullwhip effect increases when the lead-time $L$ increases, see Table \ref{tab1paper1}. 

Table \ref{tab1paper1} also shows that there is no clear relation between the autoregressive parameters and the bullwhip effect when they have different signs. In these situations the bullwhip effect may or may not exist depending on the values of $\phi_1$, $\phi_2$ and $L$, and it does not always increase when lead-time increases. These remarks confirm the results pointed out by \cite{Luong}.

%
\ctable[ caption={Bullwhip effect generated for different AR(2)
demand process.*}, label=tab1paper1, pos=!tbp, ]{lrrrr}
{\tnote[*]{SL=0.95}}
{\FL\multicolumn{1}{l}{}&\multicolumn{1}{c}{L}&\multicolumn{1}{c}{AR(c(-0.2,0.7))}&\multicolumn{1}{c}{AR(c(0.6,-0.4))}&\multicolumn{1}{c}{AR(c(0.7,0.2))}\NN
\ML &$ 1$&$0.886667$&$1.822857$&$1.315000$\NN &$
2$&$1.222133$&$1.735086$&$1.842850$\NN &$
3$&$0.970805$&$1.170277$&$2.512887$\NN &$
4$&$1.379174$&$0.917179$&$3.291280$\NN &$
5$&$1.051166$&$0.949074$&$4.141105$\NN &$
6$&$1.450366$&$1.060235$&$5.035836$\NN &$
7$&$1.097494$&$1.117111$&$5.953552$\NN &$
8$&$1.464249$&$1.103809$&$6.877221$\NN &$
9$&$1.117408$&$1.072652$&$7.793541$\NN
&$10$&$1.447477$&$1.059437$&$8.692330$\NN \LL }

In conclusion, when both first-order and second-order AR parameters are positive, the bullwhip effect exists and it increases as lead-time goes up. 
However, when the AR parameters have different signs the behaviour of the bullwhip effect is not clear. The bullwhip effect does not always exist and it is not always correct that the bullwhip effect necessarily increases when lead-time increases.

\section{Economic implications}
\label{Economic implications}
An important economic application of the use of time series methods can be seen in the safety stock level, which is the amount of 
inventory that the retailer needs to keep in order to protect himself against deviations from average demand during lead time.

Let $SS=z\sigma_d\sqrt{L}$ and $SSLT=z\hat{\sigma}^L_t$ be two safety stock measures. The former
is traditionally used in some operational research manuals and it is based on the standard deviation of the demand over $L$ periods, the latter is the safety stock as defined in (\ref{order_up}) and it is based on the standard deviation of $L$ periods forecast error.

\citealt[pp. 271,]{Chen00b} pointed out that SSLT will be greater than SS, i.e., using time series analysis, the retailer will hold more safety stock to achieve the same service level. According to the authors this is because SS captures only the uncertainty due to the random error $\epsilon$ and SSLT captures this uncertainty plus the uncertainty due to the fact that the mean demand $D_t^L$ is estimated by $\hat{D}^L_t$, in our case using the MMSE forecasting method. We show by numerical experiments that for some special cases $SSLT$ is lower than $SS$ regarding lead-time and service level. 

Using the \texttt{SCperf} function, it was verified that for ARMA and AR cases, high values on AR parameters and small values of lead-time result in lower $SSLT$.  However, in general, there is a lead-time value for which this situation is reversed. Table \ref{tab2paper1} shows the safety stock levels SS and SSLT generated by $ARMA(0.95,0.4)$ demand 
process and service level equal to $0.95$ for ten different values of lead-time, $L=1,..,10$.
For instance, for $L=2$ we have $SS=10.3$ and $SSLT=4.2$, a difference of $6$ units 
which represents a saving of $59.2\%$ over SS. Note that this difference decreases when the lead-time increases until $L=6$ where we have SSLT larger than SS.  

It is difficult to know for which value of lead-time SSLT becomes larger than SS. In general, it depends on the AR parameters of the demand.
For negative values of the AR parameters, it occurs for lower values of lead-time. Nevertheless, for the AR(2) case the AR parameters present a more complex relation with the performance of the SSLT. 
When the first-order and second-order AR parameters are positive, the pattern is the same as the AR and ARMA case, that is, SSLT 
becomes larger than SS for high values of lead-time. Moreover, when the first-order and second-order AR parameters have different signs, it is difficult to determine when the SSLT is better than SS as a measure for the safety stock level. 
%
\ctable[ caption={Bullwhip, SS and SSLT generated by
ARMA(0.95,0.4) demand process.*}, label=tab2paper1, pos=!tbp, ]{lrrrr} {\tnote[*]{SCperf(0.95,0.4,L,0.95)}} {\FL\multicolumn{1}{l}{}&\multicolumn{1}{c}{L}&\multicolumn{1}{c}{Bullwhip}&\multicolumn{1}{c}{SS}&\multicolumn{1}{c}{SSLT}\NN
\ML
&$ 1$&$1.13711$&$ 7.299$&$ 1.645$\NN
&$ 2$&$1.44321$&$10.323$&$ 4.201$\NN
&$ 3$&$1.89270$&$12.643$&$ 7.304$\NN
&$ 4$&$2.46294$&$14.598$&$10.817$\NN
&$ 5$&$3.13393$&$16.322$&$14.652$\NN
&$ 6$&$3.88802$&$17.879$&$18.745$\NN
&$ 7$&$4.70970$&$19.312$&$23.048$\NN
&$ 8$&$5.58531$&$20.645$&$27.522$\NN
&$ 9$&$6.50289$&$21.898$&$32.137$\NN
&$10$&$7.45199$&$23.082$&$36.867$\NN
\LL
}

Table \ref{tab2paper1} shows that there is a benefit resulting from the use of SSLT instead of SS as 
a measure for the safety stock level when regarding the lead-time. This benefit was verified for special demand processes where the AR parameters are high.
Moreover, if for those lead-time values where SSLT is smaller than SS, we consider the service level, it is verified that SSLT is always smaller than SS when the service level increases. 

Table \ref{tab3paper1} presents SSLT and SS generated by the same demand process for $L=1,2,3$ and ten different values 
of service level, $SL=0.9,0.91,...,0.99$. Note that when considering the service level, the difference between SS and SSLT increases for larger values of service level differently when lead-time is regarded. For instance, for $L=1$ and $SL=0.97$ we have $SS=8.35$ and $SSLT=1.88$. There is a difference of $6.47$ units which represents a saving of $77.46\%$ over SS. 

All of these facts suggest that there is a potential benefit resulting from the use of time series analysis when regarding the lead-time 
for some demand processes and, in this context, the benefit is even greater when the service level is considered. 
On the other hand, the relationship between the bullwhip effect measure and the safety stock level is more complex. Although Table \ref{tab2paper1} shows a positive relation between the bullwhip effect and the safety stock level, this relationship is not completely clear as can be seen using the SCperf function for the $AR(2)$ case when $\phi_1=-0.2$ and $\phi_2=0.7$.

In conclusion, when inventory cost and service level are of primary concern the MMSE forecast should be used since it leads in some cases to lowest safety stock level. Although the MMSE forecasting requires more computational effort, the \texttt{SCperf} function implements this method in an easy way.
\begin{table}\caption{SS and SSLT generated by
different demand processes}\label{tab3paper1}
\begin{center}
\scalebox{0.7}{
\begin{tabular}{lcccccccccc}
\hline
\multicolumn{1}{l}{\bfseries Models}&
\multicolumn{1}{c}{\bfseries Service Level}&
\multicolumn{1}{c}{\bfseries }&
\multicolumn{2}{c}{\bfseries L=1}&
\multicolumn{1}{c}{\bfseries }&
\multicolumn{2}{c}{\bfseries L=2}&
\multicolumn{1}{c}{\bfseries }&
\multicolumn{2}{c}{\bfseries L=3}
\NN 
\cline{2-2} \cline{4-5} \cline{7-8} \cline{10-11}
 &  SL &  & SS &  SSLT & &  SS &  SSLT& & SS &  SSLT\NN
\hline
&$0.90$&&$ 5.687$&$1.282$&&$ 8.043$&$3.273$&&$ 9.850$&$ 5.691$\NN
&$0.91$&&$ 5.950$&$1.341$&&$ 8.414$&$3.424$&&$10.305$&$ 5.954$\NN
&$0.92$&&$ 6.235$&$1.405$&&$ 8.818$&$3.588$&&$10.800$&$ 6.239$\NN
&$0.93$&&$ 6.549$&$1.476$&&$ 9.262$&$3.769$&&$11.343$&$ 6.553$\NN
&$0.94$&&$ 6.899$&$1.555$&&$ 9.757$&$3.971$&&$11.950$&$ 6.904$\NN
$ARMA(0.95,0.4)$&$0.95$&&$ 7.299$&$1.645$&&$10.323$&$4.201$&&$12.643$&$ 7.304$\NN
&$0.96$&&$ 7.769$&$1.751$&&$10.987$&$4.471$&&$13.456$&$ 7.774$\NN
&$0.97$&&$ 8.346$&$1.881$&&$11.803$&$4.803$&&$14.456$&$ 8.352$\NN
&$0.98$&&$ 9.114$&$2.054$&&$12.889$&$5.245$&&$15.785$&$ 9.120$\NN
&$0.99$&&$10.323$&$2.326$&&$14.599$&$5.941$&&$17.881$&$10.330$\NN
\hline
\end{tabular}
}
\end{center}\end{table}


\section{Summary}
\label{Summary}

In this paper we quantify the bullwhip effect using Zhang's result for a stationary
ARMA(p,q) demand process which admits an $MA(\infty)$ representation.
It is well known that measuring the bullwhip effect is difficult in practice. We show that using
a generalized form of this measure, the computation of this ratio is simplified if compared with traditional recursive procedures.
In some particular cases we obtain explicit formulas for this ratio.

The \texttt{SCperf} function was programmed in \textsf{R} which implements the bullwhip effect. 
We have evidenced that the use of this function makes possible accurate 
estimations of the bullwhip effect and other supply chain performance variables. We point out that no approximation is required. Moreover, we show that for certain types of demand processes the use of MMSE considered in the model leads to a significant reduction in the safety 
stock level regarding lead-time and service level. All of these observations highlight the potential economic benefits 
resulting from the use of time series analysis but it depends on the underlying demand process. For instance, if we consider an ARMA(1,1) demand processes with a high AR parameter, the use of time series techniques leads to a significant reduction in
the safety stock level but this is not the case when a low AR parameter is considered.

The \texttt{SCperf} function leads to a simple but powerful tool which gives exact analytical solutions to a
set of supply chain equations, opening up a whole new range of research opportunities.
Moreover, since the function presented in this paper is easy to use, it might be used to 
complement other managerial decision support tools. Finally, the code is given,
which makes, together with the fact that \textsf{R} is freeware, the whole research
reproducible by everyone. It may also be modified for specific tasks. 
\section*{Acknowledgements}

The author thanks Alvaro Veiga and Pat Doody for their valuable comments on earlier versions of this paper and Brigid Crowley
for a language review. This research was supported by Brazilian State Science Foundation (CAPES) grant and, in part, by the Centre for Innovation in Distributed 
Systems (CIDS - Ireland).
\appendix 
\section{}

\begin{lem} 
For a stationary ARMA(p,q) demand process, the variance of forecasting error for the lead-time demand remains constant
over time and is given by:
\begin{equation} (\hat{\sigma}^L_t)^2=Var(D_t^L-\hat{D}_t^L)=\left[1+(\sum_{j=0}^1\psi_j)^2+\cdots+(\sum_{j=0}^{L-1}\psi_j)^2\right]\sigma_{\epsilon}^2
 \quad \label{lem1}
  \end{equation}
\end{lem}
where ${\psi_j}$ satisfy (\ref{particular_psi}) and (\ref{general_psi}) and is given by (\ref{sol_psi}).
\pf Since $D_t^L=\sum_{\tau=1}^{L}d_{t+\tau}$, $\hat{D}_t^L=\sum_{\tau=1}^{L}\hat{d}_{t+\tau}$ with $\tau=1,..,L$ and 
$\hat{d}_{t+\tau}=E(d_{t+\tau}|\digamma_t)=\mu_d+\sum_{j=\tau}^{\infty}\psi_j\epsilon_{t+\tau-j}$,
the variance for the lead-time demand forecast error is 
\begin{eqnarray*}
(\hat{\sigma}^L_t)^2&=&Var[D_t^L-\hat{D}_t^L]=Var\left[\sum_{\tau=1}^{L}\left(d_{t+\tau}-\hat{d}_{t+\tau}\right)\right]\\
&=&Var\left[\sum_{\tau=1}^{L}\left(\sum_{j=0}^{\infty}\psi_j\epsilon_{t+\tau-j}-\sum_{j=\tau}^{\infty}\psi_j\epsilon_{t+\tau-j}\right)\right]\\
&=&Var\left[\sum_{\tau=1}^L\sum_{j=0}^{\tau-1}\psi_j\epsilon_{t+\tau-j}\right].
\end{eqnarray*}
By expanding the above double sum and combining the same error terms, it follows that:
$$\sum_{\tau=1}^L\sum_{j=0}^{\tau-1}\psi_j\epsilon_{t+\tau-j}=\psi_0\epsilon_{t+L}+\sum_{j=0}^1\psi_j\epsilon_{t+L-1}+\cdots+\sum_{j=0}^{L-1}\psi_j\epsilon_{t+1}$$
The independence of future error terms leads to the variance formula for lead-time demand forecast. \hfill $\square$


\section*{}
\bibliographystyle{elsarticle-harv}
\bibliography{bullwhip}
\end{document}